\documentclass[runningheads]{llncs}
\usepackage{graphicx, xcolor, amsmath, multicol, multirow, amssymb}
\usepackage{hyperref}

\begin{document}
\title{Value-Alignment Equilibrium in Multiagent Systems}
\titlerunning{Value Alignment Equilibrium in Multiagent Systems}
\author{Nieves Montes \and Carles Sierra}
\authorrunning{N. Montes \and C. Sierra}
\institute{Artificial Intelligence Research Institute, IIIA-CSIC \\
Campus UAB, Bellaterra, Spain\\
\email{\{nmontes,sierra\}@iiia.csic.es}}
\maketitle              % typeset the header of the contribution
\begin{abstract}
Value alignment has emerged in recent years as a basic principle to produce beneficial and mindful Artificial Intelligence systems. It mainly states that autonomous entities should behave in a way that is aligned with our human values. In this work, we summarize a previously developed model that considers values as preferences over states of the world and defines alignment between the governing norms and the values. We provide a use-case for this framework with the Iterated Prisoner's Dilemma model, which we use to exemplify the definitions we review. We take advantage of this use-case to introduce new concepts to be integrated with the established framework: alignment equilibrium and Pareto optimal alignment. These are inspired on the classical Nash equilibrium and Pareto optimality, but are designed to account for any value we wish to model in the system.
\end{abstract}
\section{Introduction}\label{sec:introduction}
In the last decades, research in Artificial Intelligence (AI) has been able to design and deploy increasingly complex systems, from robots and software agents, to recommendation algorithms and social networking apps \cite{luArtificialIntelligenceSurvey2019}. Given that nowadays interaction with AI systems happens on a daily basis, a new challenge arises: how to ensure that all these systems, with all their complexity and power, behave in a way that is aligned with our human values. This requirement is referred to as the Value-Alignment Problem (VAP) \cite{russellHumanCompatibleArtificial2019}, and is the focus of this work.\par

In this paper, we summarise a previously developed value alignment model \cite{sierraValueAlignmentFormal2019} motivated by the assumption that we should be able to prove that any designed system is actually complying with our values \cite{russellProvablyBeneficialArtificial2017}. It uses \textit{norms} as the essential tools to supervise and limit autonomous agents' behaviour \cite{mahmoud2014}. More importantly, it provides a precise definition of what it means for a norm to be aligned with a given value. We then provide a use-case based on the benchmark Iterated Prisoner's Dilemma game \cite{axelrod2006}. Inspired by the concepts of Nash equilibria and Pareto optimality in game theory, we introduce new concepts of value equilibria and optimality that build on top of the formal framework in \cite{sierraValueAlignmentFormal2019}.

\section{Value alignment model}\label{sec:theory}
\subsection{Revision of our background formal model}\label{subsec:formal_model}
In order to introduce the necessary background, we provide a summary of the value alignment model that constitutes the starting point of this paper \cite{sierraValueAlignmentFormal2019}. Its main underlying assumption is a \textit{consequentialist view of values} \cite{sullivanEthicalTheories2002}, which expresses that the worthiness of any value is entirely determined by the outcomes of the actions that it motivates. Values, then, can serve as numerical quantifiers to assess how (un)desirable is a state in the world. In particular, we can use values to compare any two states and decide which of the two is preferred.\par

In the reviewed framework, the common conception of the world as a labelled transition system \cite{gorrieriLabeledTransitionSystems2017} is adopted. The world, then, is composed of a set of states $\mathcal{S}$, a set of actions $\mathcal{A}$ and a set of transitions $\mathcal{T\subseteq \mathcal{S}\times\mathcal{A}\times\mathcal{S}}$. We refer to any transition $\left(s, a, s'\right)\in\mathcal{T}$ with the notation $s\xrightarrow{a}s'$.\par

Values are conceived as mental constructs \cite{miceliCognitiveApproachValues1989} that allow agents to decide which state of the world they prefer, according to their most prioritised values. This consideration motivates the following definition:

\begin{definition}\label{def:preferences}
{\normalfont (from \cite{sierraValueAlignmentFormal2019})} A value-based preference $\mathsf{Prf}$ is a function over pairs of states that indicates how much preferred is one state over another in light of a particular value: $\mathsf{Prf}:\mathcal{S}\times\mathcal{S}\times G \times V\rightarrow [-1,1]$, where $G$ is the set of agents and $V$ is the set of values. The notation $\mathsf{Prf}_v^\alpha(s,s')$ indicates how much does agent $\alpha$ prefer state $s'$ over state $s$ with respect to value $v$.
\end{definition}

Value-based preferences, then, are bounded functions between -1 and +1. Positive (negative) preference indicates the post-transition state is more (less) desirable than the pre-transition state with respect to a specific value $v$. Preferences equal to 0 indicate that both states are identically preferred.\par

Values are held at the agent level. However, very rarely do agents' belief systems consist of a single value \cite{schwartz2012}. Aggregations over subsets of values and/or agents are also considered in \cite{sierraValueAlignmentFormal2019}. However, we will not be employing any aggregation functions in this work. The interested reader is referred to the original paper for further details.\par

Now we have a clear view of the role of values when it comes to evaluating states. However, the states that can arise when letting a multiagent system evolve are dependent upon the \textit{norms} in place, since these are the constructs which govern behaviour and therefore limit the actions that can be taken. Thus, value alignment is conceived as the alignment of a norm (or a set of norms) with respect to a value that is held in high regard. When a set of norms $N$ is incorporated into the world $(\mathcal{S}, \mathcal{A}, \mathcal{T})$, it is modified into a new, \textit{normative} world $(\mathcal{S}, \mathcal{A}, N, \mathcal{T}_N)$ \cite{andrighettoNormativeMultiAgentSystems2013}, where $\mathcal{T}_N\subseteq\mathcal{T}$ is the subset of all the original transitions allowed by $N$.\par

To evaluate how well or badly aligned is a norm $n\in N$, the transitions that can happen when this norm is enforced have to be evaluated. A norm is positively (negatively) aligned if it gives rise to transitions that move the system towards more (less) preferred states. However, beyond single transitions, the long-term evolution of the world under the norms in place should be considered. This necessity motivates the following definition:

\begin{definition}\label{def:path}
{\normalfont (from \cite{sierraValueAlignmentFormal2019})} A {\normalfont path} $p$ in the world $(\mathcal{S}, \mathcal{A}, \mathcal{T})$ is a finite sequence of consecutive transitions $\{s_0\xrightarrow{a_0}s_1, s_1\xrightarrow{a_1}s_2, ..., s_i\xrightarrow{a_i}s_{i+1},...,s_\mathit{f}\xrightarrow{a_{\mathit{f}-1}}s_\mathit{f}\}$, where $p_F[i]=p_I[i+1]$, and $p_I[i]$ and $p_F[i]$ denote the pre- and post-transition states of the $i$-th transition.
\end{definition}

Paths are used to evaluate preferences over consecutive transitions, and support the formal definition of value alignment:

\begin{definition}\label{def:alignment}
{\normalfont (from \cite{sierraValueAlignmentFormal2019})} The degree of alignment of norm $n\in N$ with respect to value $v\in V$ in the world $\left(\mathcal{S}, \mathcal{A}, \mathcal{T}\right)$ for agent $\alpha\in G$ is defined as the accumulated preference over all the paths in the normative world that results from implementing such norm:
\begin{equation}\label{eq:alignment_full}
    \mathsf{Algn}_{n, v}^\alpha \left(\mathcal{S}, \mathcal{A}, \mathcal{T}\right) = \frac{\sum\limits_{p\in \text{paths}} \sum\limits_{d\in [1, |p|]} \mathsf{Prf}_v^\alpha (p_I[d], p_F[d]) }{\sum\limits_{p\in \text{paths}} |p|}
\end{equation}
where {\normalfont paths} is the set of all paths in the normative world $\left(\mathcal{S}, \mathcal{A}, \{n\}, \mathcal{T}_n\right)$, and $|p|$ corresponds to the cardinality of $p$, {\em i.e.} the number of transitions in the path.
\end{definition}

Note that we exclude the possibility of infinite transition systems by considering all paths to be finite.\par

In  this approach, the same exact weight is given to every single transition. Other suggestions are conceivable; for example one may want to consider whether preferences remain approximately stable along the paths or, conversely, there are large surges or sinks. In other fields, it is typical to consider a discount parameter that reduces the weight of transitions that happen into the distant future \cite{schwartzMultiAgentMachineLearning2014}. We acknowledge the existence of alternative approaches, but leave their exploration for future work.\par

Another issue to note with definition \ref{def:alignment} is that its notation indicates the alignment for a single agent $\alpha$ with respect to a single value $v$, since it takes the individual preferences with respect to that one value. However, as previously mentioned, preferences can be aggregated over agents and/or sets of values. Using such aggregated preferences would, consequently, result in alignment for sets of agents and/or with respect to sets of values.\par

Equation \eqref{eq:alignment_full} states that alignment should take into account all possible transitions in the normative world. In general, this approach is not efficient, and to solve this issue, Monte Carlo sampling over all possible paths is recommended. Additionally, it is also advisable to keep the length of the paths fixed. These modifications lead to the following reformulation for the alignment:
\begin{equation}\label{eq:alignment_changed}
    \mathsf{Algn}_{n, v}^\alpha \left(\mathcal{S}, \mathcal{A}, \mathcal{T}\right) = \frac{\sum\limits_{p\in \text{paths}} \sum\limits_{d\in[1,l]} \mathsf{Prf}_v^\alpha (p_I[d], p_F[d]) }{x\times l}
\end{equation}
where $l$ is the number of transitions in all paths, and $x$ is the number of sampled paths.\par

In summary, the approach proposed in \cite{sierraValueAlignmentFormal2019} provides a formal model to numerically quantify how compliant is a certain normative world designed towards some particular value. Differently to other works related to value alignment \cite{atkinson2016,cranefield2017}, such a model is separate from the decision-making process of the participating agents and their respective goals, and can hence be applied to any kind of agent and social space architecture.

\subsection{Value-alignment solution concepts}\label{subsec:solution_concepts}
In this work, we model agents' interactions as normal-form games. In game theory, a stage game refers to each of the identical rounds played in one iteration. The Nash equilibrium is then defined as the set of players' actions such that no player can obtain a higher profit by unilaterally deviating from it, given the actions of all other players are fixed \cite[Chapter 2]{osborneCourseGameTheory2012}. In formal terms, Nash equilibria correspond to action profiles $(a^*_1,...,a^*_{|G|})$ such that, for all agents $i\in G$, it holds that:
\begin{equation}\label{eq:nash_eq}
r_i(a^*_1,...,a^*_{i-1}, a^*_i, a^*_{i+1},...,a^*_{|G|}) \geq r_i(a^*_1,...,a^*_{i-1}, a_i, a^*_{i+1},...,a^*_{|G|})
\end{equation}

Nash equilibria may not represent the best option for players in terms of individual revenues, but profiles of joint actions for which no player has an incentive to unilaterally deviate. This solution concept represents \textit{status quo} positions that persist despite not necessarily being the best solutions from a social perspective. In other terms, the Nash equilibria do not necessarily correspond with Pareto optimal outcomes \cite{pardalosParetoOptimalityGame2008a}, as in the case of the classical Prisoner's Dilemma game.\par

The concept of stage game Nash equilibrium is not directly applicable in our theoretical framework for two main reasons. First, in general, agents interact repeatedly. And second, we are interested in the alignment with respect to values, not in the game rewards themselves, even if alignments may be computed from rewards. However, Nash-like equilibria situations can be identified when the alignment satisfies an adapted version of equation \eqref{eq:nash_eq}. When considering this possibility, the argument $a^*_i$ should not be identified as actions taken by agents in a single round, but rather by the strategies that individuals follow for the whole duration of the game. We conceive individual strategies as separate from norms. While strategies are part of the agents' internal decision process, we understand norms to be externally imposed constraints on the system, whose definition is the responsibility of the model's designer and beyond the control of the participating agents.\par

Therefore, given a set of norms $N$ governing a multiagent system, agents adopt a particular strategy to play the game. Strategies are functions that take into account the past history of the game to return an action to be performed next \cite{osborneCourseGameTheory2012}. The set of all strategies being played is $E=\{e_1, e_2,...,e_{|G|}\}$, where $e_i$ is the strategy followed by the $i$-th agent. Then, we can make the following definition:
\begin{definition}\label{def:algn_equilibrium}
Given a normative system $N$ and an alignment function with respect to the value of choice for agent $i$ $\mathsf{Algn}^i_v$, the {\em alignment equilibrium} is defined as the tuple of all individual strategies $\left(e^*_1,..., e^*_{|G|}\right)$ such that, for all agents $i\in G$, it holds that:
\begin{equation}\label{eq:nash_algn}
\mathsf{Algn}^i_{\left(e^*_1,..., e^*_{i-1}, e^*_{i}, e^*_{i+1},..., e^*_{|G|}\right), v} \geq \mathsf{Algn}^i_{\left(e^*_1,..., e^*_{i-1}, e_{i}, e^*_{i+1},..., e^*_{|G|}\right), v}
\end{equation}
\end{definition}

Note that the alignment equilibria depend on the alignment function of choice and the norms constraining the system. Another point to note is that, despite equation \eqref{eq:nash_algn} is stated in terms of alignment with respect to individual agents, a natural extension arises when applied to subsets of agents or even aggregated for the whole society. Additionally, it can also be extended to sets of values.\par

Another important point is that, unlike the classical Nash equilibrium, whose existence is guaranteed for games played by a finite number of agents following mixed strategies \cite{NashTheoremGame2002}, the properties of equation \eqref{eq:nash_algn} have not been explored to the point of establishing conditions for existence. This is left as future work.\par

It is worth trying to understand the motivation behind equation \eqref{eq:nash_algn}. Classical game theory takes the view of trying to maximise one's own reward while minimising risks. The classical Nash equilibrium for a stage game is a possible response to this approach for a game that is played once. Our definition of alignment equilibrium, then, generalises the classical Nash equilibrium to account for games and situations that are presented repeatedly.\par

More importantly, equation \eqref{eq:nash_algn} allows to examine a game from a different perspective other than that of the individual payoffs $r_i$, as we can analyse the alignment equilibria with respect to as many values as alignment functions we are able to come up with. Moreover, agents may consider different preferences or may take into account different variables to compute their preferences, and thus equation \eqref{eq:nash_algn} need not be symmetric with respect to agents. By considering the particular instance where a game is played for a path of length one (a single round) and using the actual rewards as preferences, the classical concept of Nash equilibrium can be recovered.\par

Previously, we have mentioned the concept of Pareto optimality \cite{pardalosParetoOptimalityGame2008a}. This solution concept from game theory can also be adapted to the value alignment framework. Classically, an action profile $(a^*_1,...,a^*_{|G|})$ is said to lead to a \textit{Pareto optimal outcome} if there is no other profile $(a_1,...,a_{|G|})$ such that:
\begin{align}
    r_j\left(a_1,...,a_{|G|}\right) &> r_j\left(a^*_1,...,a^*_{|G|}\right)\;\text{for at least one agent }j\in G,\;\text{and}\\
    r_i\left(a_1,...,a_{|G|}\right) &\geq r_i\left(a^*_1,...,a^*_{|G|}\right)\;\text{for all agents } i\in G
\end{align}
In other words, in classical game theory a Pareto optimal outcome is one where we cannot improve anyone's reward without damaging someone else's.\par

In a similar fashion to our extension from the classical Nash equilibrium to the alignment equilibrium, we can adapt Pareto optimality to our value alignment framework. Again, we need not consider actions for a single transition, but rather strategies adopted during the whole duration of the game:
\begin{definition}
Given a normative system $N$ and an alignment function with respect to the value of choice for agent $i$ $\mathsf{Algn}^i_v$, a tuple of individual strategies $\left(e^*_1,..., e^*_{|G|}\right)$ is said to lead to a {\normalfont Pareto optimal alignment} if there is no other tuple of individual strategies $\left(e_1,..., e_{|G|}\right)$ such that:
\begin{align}\label{eq:pareto_algn}
\mathsf{Algn}^j_{\left(e_1,...,e_{|G|}\right), v} &> \mathsf{Algn}^j_{\left(e^*_1,...,e^*_{|G|}\right), v}\;\text{\normalfont for at least one agent }j\in G,\;\text{\normalfont and}\\
\mathsf{Algn}^i_{\left(e_1,...,e_{|G|}\right), v} &\geq \mathsf{Algn}^i_{\left(e^*_1,...,e^*_{|G|}\right), v}\;\text{\normalfont for all agents }i\in G
\end{align}
\end{definition}

Therefore, a Pareto optimal alignment corresponds to a situation where no agent can improve its alignment without hurting one or several other agent alignments. The novel concept of Pareto optimal alignment allows us to make the same generalisations than those when moving from the classical Nash equilibrium to the alignment equilibrium. They have been explicitly described previously.\par

\section{Use-case: Two-Agent Iterated Prisoner's Dilemma}\label{sec:use_case}
Now, we consider a very simple two-agent system modelled after the benchmark Iterated Prisoner's Dilemma (IPD) game \cite{axelrod2006}, and we use it to illustrate the previously introduced concepts.\par

\begin{table}[ht]
	\centering
	\caption{Outcome matrix for the Prisoner's Dilemma game.}
	\label{tab:PD_matrix}
	\begin{tabular}{|c|c|c|c|}
		\cline{3-4} \multicolumn{2}{c|}{} & \multicolumn{2}{|c|}{$\beta$} \\
		\cline{3-4} \multicolumn{2}{c|}{} & Cooperate & Defect \\
		\hline \multirow{2}{*}{$\alpha$} & Cooperate & $(6,6)$ & $(0,9)$ \\
		\cline{2-4}  & Defect & $(9,0)$ & $(3,3)$ \\ \hline
	\end{tabular}
\end{table}

In formal terms, this model consists of an agent set with two members, $G=\{\alpha, \beta\}$; the set of individual actions available to each of them is $\mathcal{A}_i=\{\text{cooperate}, \text{defect}\}$. The transitions between states are characterised by the joint actions that $\alpha$ and $\beta$ take, hence the set of actions is $\mathcal{A}=\mathcal{A}_i^2=\{(a_\alpha, a_\beta)\}$. The tuple of individual actions $(a_\alpha, a_\beta)$ determines the outcome of the transition. Such outcomes are tied to rewards agents receive, $r_i(a_\alpha, a_\beta),\space \forall i \in G$, via the outcome matrix displayed in Table \ref{tab:PD_matrix}.\par

\begin{figure}[ht]
    \centering
    \includegraphics[width=0.4\linewidth]{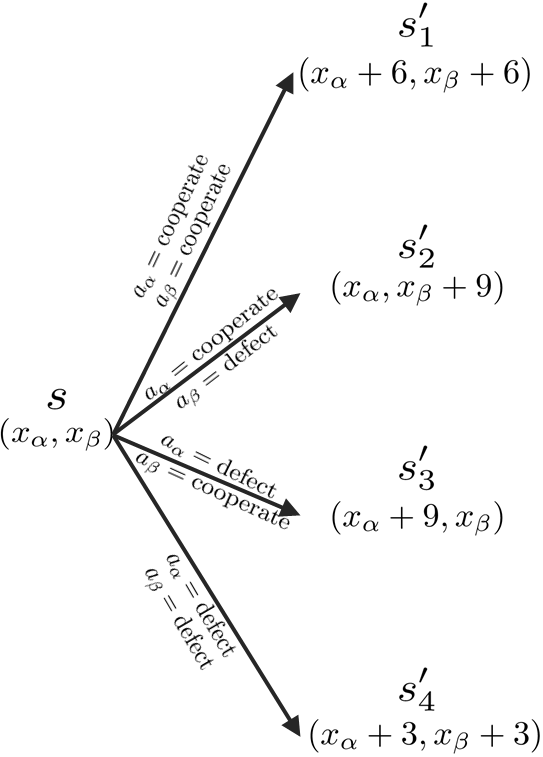}
    \caption{Labelled transition system representing a single transition of the 2A-IPD model.}
    \label{fig:transitions}
\end{figure}

In this model, the game is played in an iterated way. We characterise states by the agents' wealth, which is defined as the accumulated rewards received since the game started. The states of the world correspond to the tuple of agents' wealth, noted by $(x_\alpha, x_\beta)$. Initially, both agents start with $(x_\alpha, x_\beta)=(0,0)$. After agents have chosen their actions $(a_\alpha, a_\beta)$ and the rewards $(r_\alpha, r_\beta)$ are determined, the transition $s \rightarrow s'$ is completed by updating the values for agents' wealth.\par

We formally lay out the main features of our model in the following terms:

\begin{definition}\label{ch2:def:2A-IPD}
We define the {\em Two-Agent Iterative Prisoner's Dilemma (2A-IPD)} model as a tuple $(G, \mathcal{S}, \mathcal{A}, \mathcal{T})$, where:
\begin{itemize}
    \item $G=\{\alpha, \beta\}$ is a set of two agents.
    \item $\mathcal{S} = \{\left(x_\alpha,x_\beta\right)\}$ is the set of states, composed by the tuples of all possible combinations of agents' wealth.
    \item $\mathcal{A}=\{(a_\alpha, a_\beta)\}$, where $a_i \in \{\text{cooperate, defect}\}$, is the set of joint actions.
    \item $\mathcal{T}: \mathcal{S}\times\mathcal{A}\times\mathcal{S}$ is the set of transitions which relate the current state $s = \left(x_\alpha, x_\beta\right)$ and joint actions $\left(a_\alpha, a_\beta\right)$ to the next state $s' = (x'_\alpha, x'_\beta)$:
    \begin{equation}\label{ch2:eq:transition}
    \forall t \in \mathcal{T},\quad t = \left(s, (a_\alpha, a_\beta), s'\right) \quad\text{such that}\quad
    \begin{cases}
    x'_\alpha =& x_\alpha + r_\alpha(a_\alpha, a_\beta)\\
    x'_\beta =& x_\beta + r_\beta(a_\alpha, a_\beta)
    \end{cases}
    \end{equation}
    where $r_i(a_\alpha, a_\beta)$ are the rewards given by the outcome matrix (Table \ref{tab:PD_matrix}).
    \end{itemize}
\end{definition}

A labelled transition system representation for one transition of this model is displayed in Fig. \ref{fig:transitions}.

\subsection{Preference functions}\label{subsec:PD_preferences}

Within this model, we wish to quantify the alignment of different agent behaviours with respect to two apparently opposed values: {\em equality} and {\em personal gain}. In \cite{sierraValueAlignmentFormal2019} some preferences related to these values were already formulated. However, we introduce new ones here that are less sensitive to the numerical choice for the rewards in Table \ref{tab:PD_matrix}.\par

In order to quantify alignment with respect to equality, we will assess it in each state by making use of the well-known Gini Index (GI) \cite{cowellMeasuringInequality2009}. In our two-agent model, the GI is computed for state $s$ by taking the values of agents' wealth at that particular point in the repeated game, $(x_\alpha, x_\beta)$. For $|G|=2$, the Gini Index for our system becomes:
\begin{equation}\label{eq:GI_2agents}
\text{GI}(s) = \frac{|x_\alpha - x_\beta|}{2\left(x_\alpha + x_\beta\right)}
\end{equation}

The lower bound for the GI is always 0, indicating perfect equality among all participants. In the case of two agents, the maximum possible value is $\tfrac{1}{2}$ \cite{belluInequalityAnalysisGini2006}. Then, in order to map perfect equality ($\text{GI}=0$) to maximum preference (+1) and perfect inequality ($\text{GI}=\tfrac{1}{2}$) to minimum preference (-1), the interval for the Gini Index $\left[0,\tfrac{1}{2}\right]$ is linearly mapped to the interval of definition of preferences $\left[-1,1\right]$. This transformation results in the following definition for the preference function over the value {\em equality}:
\begin{equation}\label{eq:pref_equality}
\mathsf{Prf}_{\text{equality}}^i(s, s') = 1-4\cdot \text{GI}(s') = 1 - 2\cdot\frac{|x'_\alpha - x'_\beta|}{x'_\alpha + x'_\beta}
\end{equation}
where $i=\alpha,\beta$.\par

There are two important points to be noted about this preference function. First, it is numerically equivalent for both agents, since $x'_\alpha$ and $x'_\beta$ are interchangeable. It is an intuitive property of the preference with respect to the value \textit{equality} that it should be indeed identical for both agents. Second, $\mathsf{Prf}_{\text{equality}}^i(s, s')$ is a function of only the properties of the system in the post-transition state $s'$. Hence, states with high (low) equality are (not) preferred regardless of the parity in the previous state $s$.\par

The latter property is not a requirement of preference functions. Other formulae could be devised that depend on the increase/decrease of the Gini Index or some other indicator. Since our model starts off from a very peculiar position of perfect equality, we have considered that it is more helpful to monitor the eventual disparity that may arise as the game proceeds, rather than comparing consecutive states.\par

The other value under consideration, personal gain, is quantified through the following preference function:
\begin{equation}\label{eq:pref_gain}
\mathsf{Prf}_\text{gain}^i (s, s') = \begin{cases}
-1 & \text{if } x'_i - x_i = 0 \\
- \frac{1}{3} & \text{if } x'_i - x_i = 3 \\
\frac{1}{3} & \text{if } x'_i - x_i = 6 \\
1 & \text{if } x'_i - x_i = 9
\end{cases}
\end{equation}
According to this definition, states are preferred with respect to personal gain by ranking the possible rewards in any given transition and mapping them to equally spaced points in the preference interval $[-1, 1]$. This choice is made to reflect the greediness of this value, since the preference is only dependent on immediate gains, regardless of how well off the agent may already be or the circumstances of her peer.\par

Now, we have the preference functions to evaluate the alignment of the system for the two agents. In order to compute the alignment from preferences, equation \eqref{eq:alignment_changed} is employed, with $x=10,000$ sampled paths of length $l=10$ (the number of game iterations).\par

\subsection{Individual strategies}\label{subsec:strategies}
Now that we have encoded values into two different preference functions, our purpose, then, is to find which agent strategies in the 2A-IPD model result in alignment equilibrium and Pareto optimal alignment positions with respect to equality and personal gain. The set of all possible strategies for a single agent is a vast space of logical formulae, dictating whether past actions should be taken into account to decide the future action, how far back to look and many other considerations. In order to work with the alignment equilibrium without the burden of considering \textit{all} possible strategies, we restrict ourselves to two subspaces of possible strategy profiles, from which we will determine the alignment equilibria and Pareto optimal alignments:

\begin{enumerate}
	\item \textit{Random-action profiles:} Both $\alpha$ and $\beta$ choose at each round the action to take randomly, according to independent and fixed probabilities of cooperation. These are analogous to mixed strategies in game theory, where actions are taken based on a fixed probability.
	\item \textit{Heterogeneous profiles:} $\beta$ cooperates randomly following a fixed probability. $\alpha$, in contrast, either cooperates or defects with probability 0.5 in the first round. In subsequent rounds, it follows one of these strategies:
	\begin{itemize}
		\item \textit{Tit-for-tat:} $\alpha$'s action in the current round is $\beta$'s action in the previous round.
		\item \textit{Mostly cooperate:} $\alpha$ defects if in the previous round both agents defected. Otherwise, it cooperates.
		\item \textit{Mostly defect:} $\alpha$ cooperates if in the previous round both players cooperated. Otherwise, it defects.
	\end{itemize}
	These are some of the most common strategies in non-cooperative game theory. We set that only one agent follows them in order not to condition the outcome of all transitions on the result of the first one.\par
\end{enumerate}
Note that all of the strategies presented here are included in the set of reactive strategies \cite{nowak1992}, where behaviour is only dependent on the opponent's immediate past action.

\section{Results}\label{sec:results}
\begin{figure}[ht]
	\centering
	\includegraphics[width=0.5\linewidth]{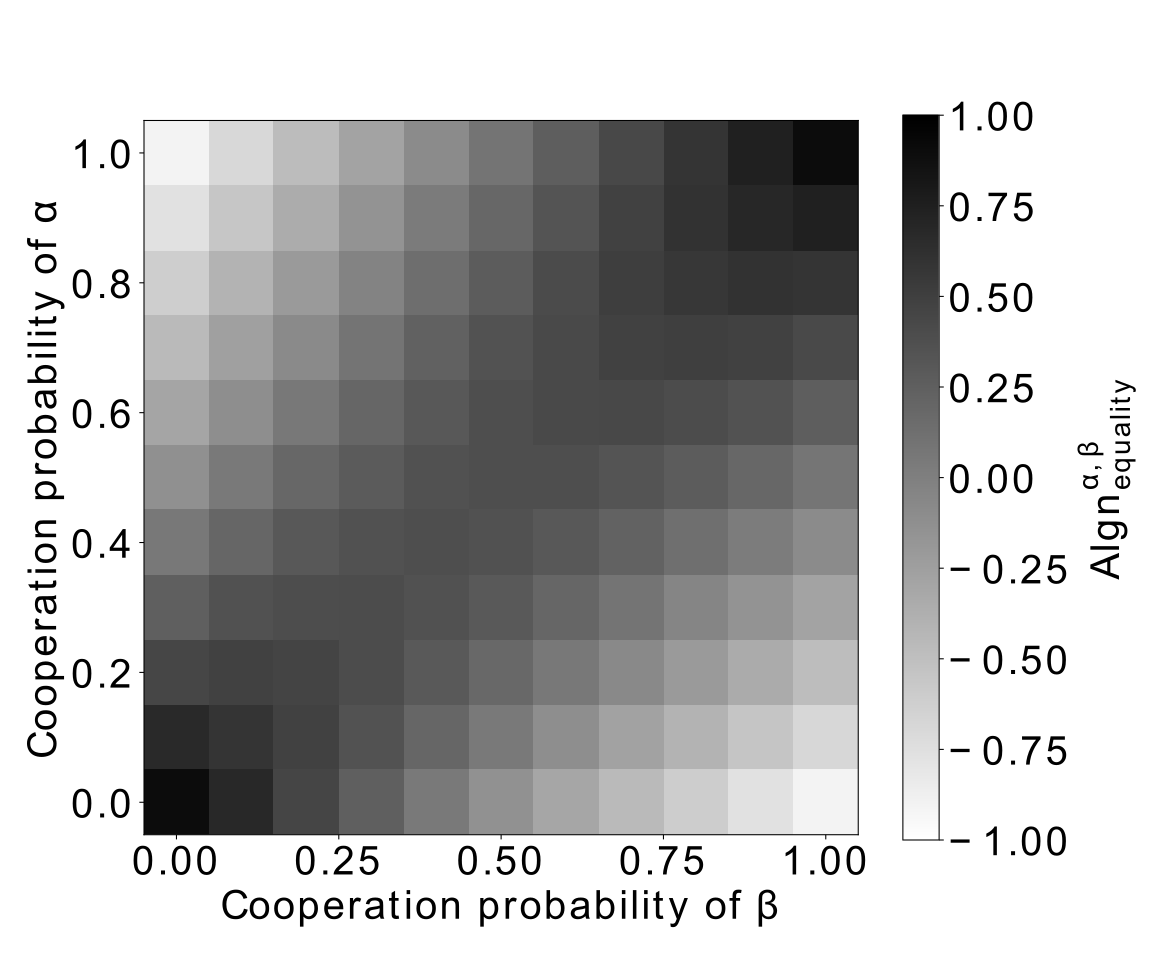}
	\caption{Alignment of agents $\alpha$ and $\beta$ with respect to value {\em equality}, equation \eqref{eq:pref_equality}, under random-action profiles.}
	\label{fig:random_pd_algn_eq}
\end{figure}

\begin{figure}[ht]
	\centering
	\includegraphics[width=\linewidth]{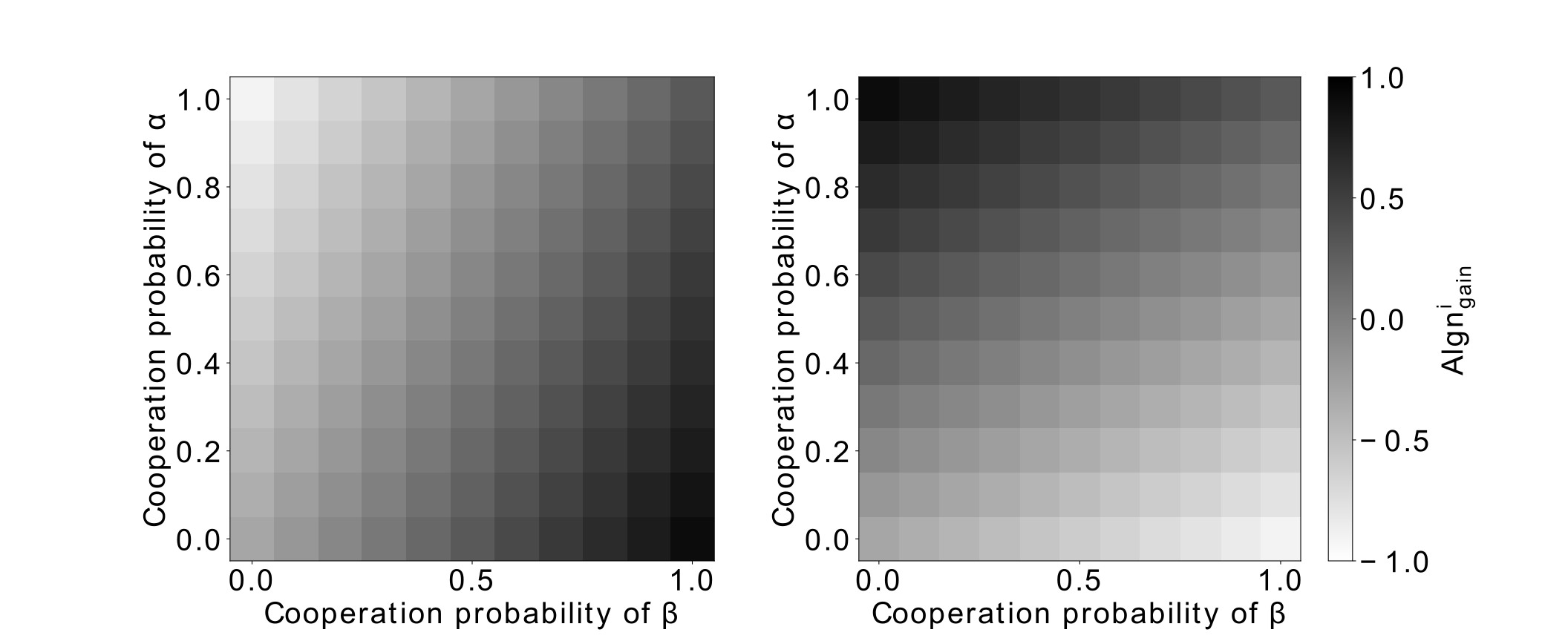}
	\caption{Alignment of agent $\alpha$ (left) and $\beta$ (right) with respect to value {\em personal gain}, equation \eqref{eq:pref_gain}, under random-action profiles. Note that $\mathsf{Algn}^\alpha_{\text{gain}}$ and $\mathsf{Algn}^\beta_{\text{gain}}$ are mutually transposed.}
	\label{fig:random_pd_algn_gain}
\end{figure}

First, the results for the alignment under random-action profiles with respect to equality and personal gain are presented in Figs. \ref{fig:random_pd_algn_eq} and \ref{fig:random_pd_algn_gain} respectively. They are plotted as a function of the cooperation probabilities of both agents, which specify concrete instances of the individual random strategies $e_\alpha$ and $e_\beta$.\par

According to the definition of alignment equilibrium in equation \eqref{eq:nash_algn}, in Fig. \ref{fig:random_pd_algn_eq} there is an infinite number of alignment equilibria corresponding to strategies satisfying $e_\alpha = e_\beta$ (here we only plot 10$\times$10 strategy pairs). To see this, if we consider that the cooperation probability of agent $\beta$ is $e_\beta$, then $\alpha$'s alignment is maximal when $e_\alpha$ is equal to $e_\beta$, and $\alpha$ has no incentive to deviate. Similarly if we fix the cooperation probability of agent $\alpha$, given the symmetry of equation \eqref{eq:pref_equality}. Therefore, an infinite number of alignment equilibria are found at $e^*_\alpha=e^*_\beta$.\par

It is unsurprising that the \textit{status quo} with respect to equality corresponds to both agents having the same cooperation probability. It is worth noting that this result does not encourage any agent to increase its cooperation probability to enhance equality, but rather to act similarly to her peer.\par

The Pareto optimal alignment with respect to equality is found when either both agents always cooperate or always defect, since they always receive identical rewards and the Gini Index is kept to zero. So, actually, alignment equilibrium strategies with respect to equality do include Pareto optimal strategies.\par

As for alignment with respect to gain in Fig. \ref{fig:random_pd_algn_gain}, the maximum alignment for any agent is obtained when she defects and the other player follows a complete cooperation strategy. The alignment equilibrium, again considering only random-action profiles, is found at the position where both agents never cooperate. This is a consequence of the single Nash equilibrium of the stage game at the position (defect, defect), in combination with the definition of preference with respect to personal gain in equation \eqref{eq:pref_gain} being directly related to the individual gains for a single round.\par

Differently from the results obtained for value {\em equality} where equilibrium positions were Pareto optimal, for value {\em personal gain} this is not the case. Starting from constant mutual defection, both agents could improve their alignment by turning to constant mutual cooperation instead, since they would attain larger gains at each round. This result is also due to the direct relationship between preferences with respect to personal gain and the actual rewards in each round, see equation \eqref{eq:pref_gain}.\par

\begin{table}[b]
    \centering
    \caption{Summary of the results obtained for two randomly-behaving players, highlighting the position of alignment equilibrium strategies and their Pareto optimality.}
    \begin{tabular}{|c|c|c|c|}
        \hline \textbf{Value for} $\mathbf{\alpha}$ & \textbf{Value for} $\mathbf{\beta}$ & \textbf{Alignment equilibrium} & \textbf{Pareto optimal} \\
        \hline Equality & Equality & $e_\alpha^*=e_\beta^*$ & Included \\
        \hline Personal gain & Personal gain & $e_\alpha^*=e_\beta^*=0$ & No \\
        \hline Personal gain & Equality & $e_\alpha^*=e_\beta^*=0$ & No \\
        \hline
    \end{tabular}
    \label{tab:random_profiles_summary}
\end{table}

This pair of strategies (both agents always defecting for all repetitions of the game) is the alignment equilibrium when both agents' alignments are computed with respect to value  {\em personal gain}. An interesting observation is that this is also the alignment equilibrium when one player's alignment is computed with respect to equality, and the other with respect to gain. Let's consider the case where $\alpha$'s alignment is computed with respect to gain, Fig. \ref{fig:random_pd_algn_gain} left, and $\beta$'s alignment is computed with respect to equality, Fig. \ref{fig:random_pd_algn_eq}.\par

Regardless of $\beta$'s actions, $\alpha$'s alignment is always maximised under null cooperation probability, so the preferred strategy to take is to defect in every round. Then, $\beta$ maximises her alignment with respect to equality by imitating $\alpha$, that is, always defecting as well. The result is that the only alignment equilibrium satisfying equation \eqref{eq:nash_algn} for random-actions profiles is found at $e_\alpha^* = e_\beta^* = 0$. It is worth noting that the calculation for the alignment is specific to each agent in order to reflect their different priority values. Again, this alignment is not Pareto optimal. If both agents switched to $e_\alpha = e_\beta = 1$, $\alpha$ could enhance her alignment with respect to personal gain while maintaining $\beta$'s alignment with respect to equality at 1.\par

A summary of the results presented so far for random-action profiles is displayed in Table \ref{tab:random_profiles_summary}.\par

\begin{figure}[t]
	\centering
	\includegraphics[width=0.65\linewidth]{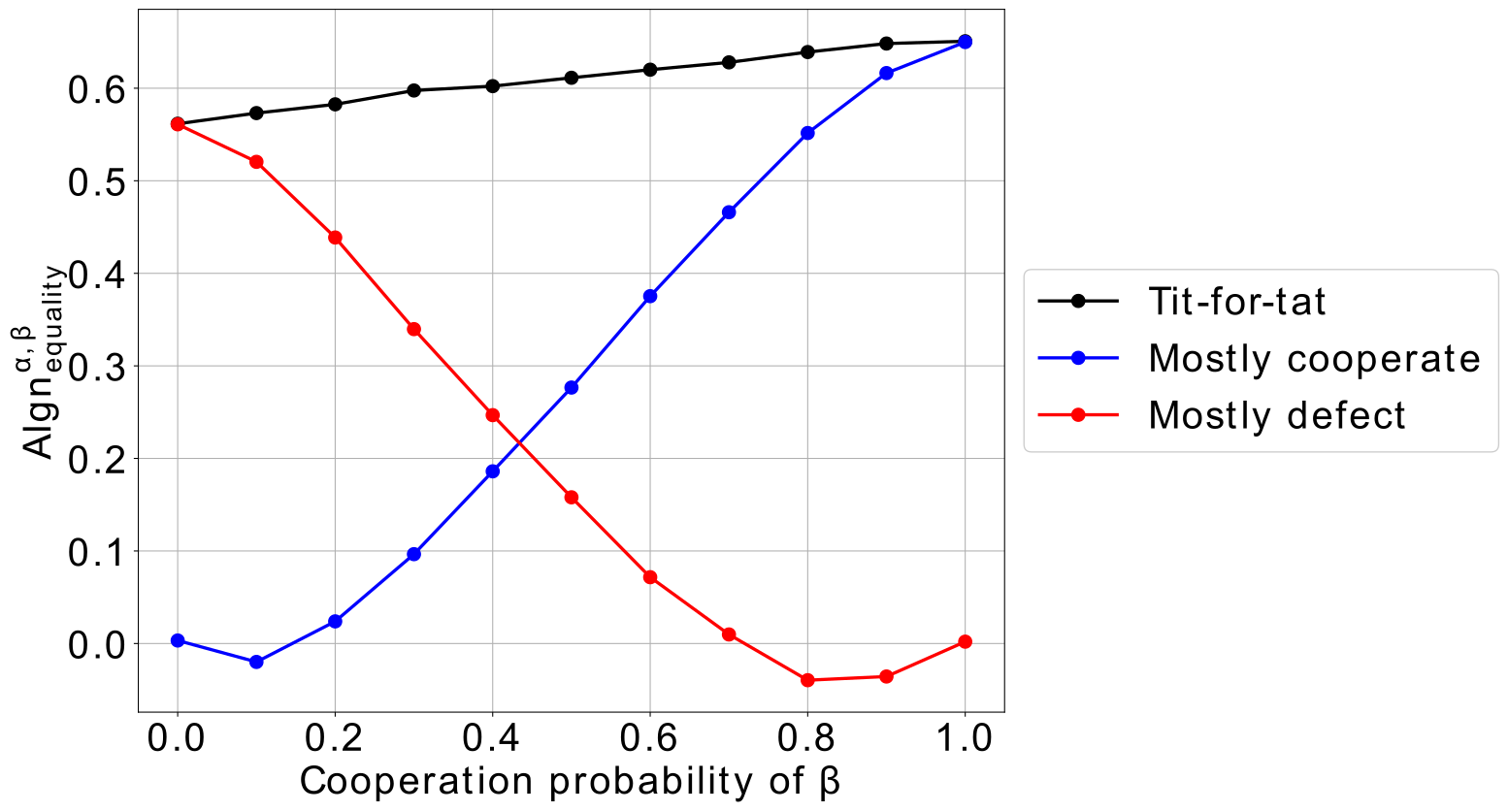}
	\caption{Alignment of agents $\alpha$ and $\beta$ with respect to value {\em equality}, equation \eqref{eq:pref_equality}, under heterogeneous profiles, depending on the strategy followed by player $\alpha$.}
	\label{fig:alpha_strat_algn_eq}
\end{figure}

\begin{figure}[t]
    \centering
	\includegraphics[width=\linewidth]{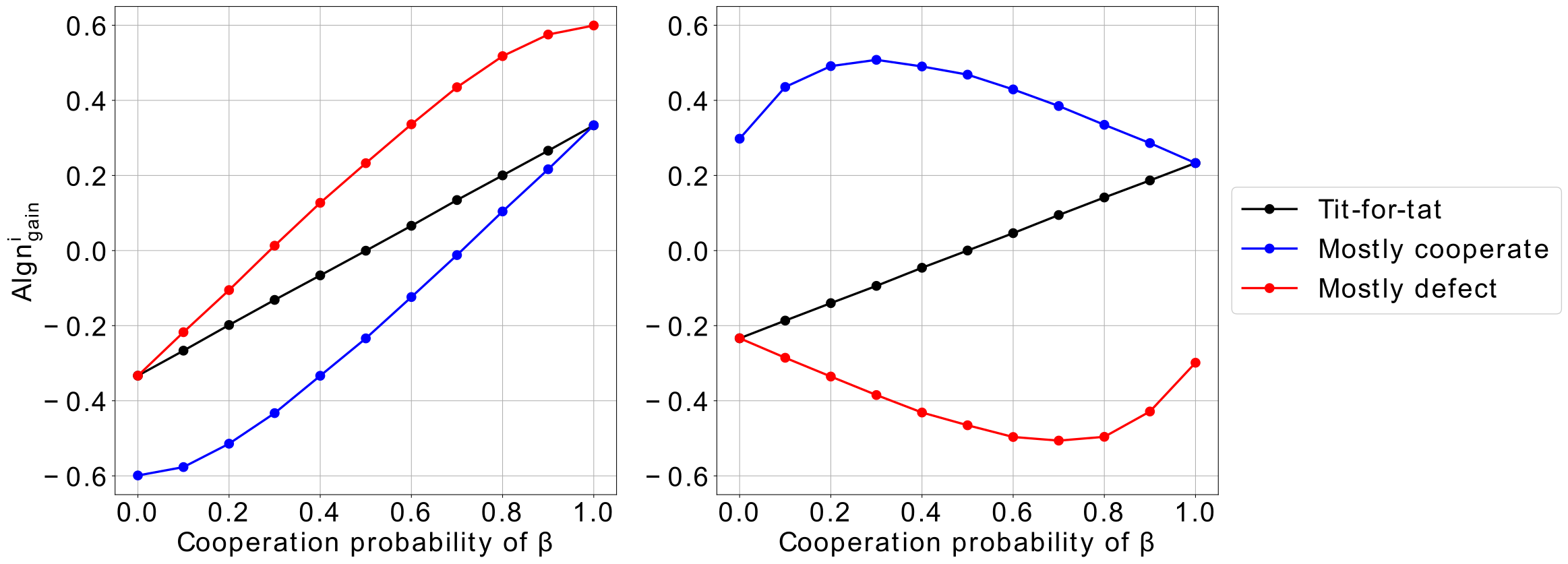}
	\caption{Alignment of agent $\alpha$ (left) and $\beta$ (right) with respect to value {\em personal gain}, equation \eqref{eq:pref_gain}, under heterogeneous profiles, according to the strategy followed by player $\alpha$.}
    \label{fig:alpha_strat_algn_gain}
\end{figure}

Now, we look into the alignment with respect to a single value under heterogeneous profiles. The alignment for both players is plotted as a function of the cooperation probability of $\beta$, under the various possible strategies for $\alpha$, for value {\em equality} in Fig. \ref{fig:alpha_strat_algn_eq} and for value {\em personal gain} in Fig. \ref{fig:alpha_strat_algn_gain}.\par

First, we focus our attention on the alignment with respect to equality in Fig. \ref{fig:alpha_strat_algn_eq}. In this case, the three strategies result in very distinct trends. \textit{Tit-for-tat}, equivalent to imitating the opponent, leads to a very stable alignment with respect to equality, independently of the extend of collaboration of $\beta$. The other two strategies, \textit{mostly cooperate} and \textit{mostly defect}, result in alignments that are strongly dependent on $\beta$'s cooperation probability. When $\beta$ defects often (cooperation $\sim 0$), \textit{mostly defect} is the preferred individual strategy for $\alpha$. In contrast, when $\beta$ cooperates often (cooperation $\sim 1$), \textit{mostly cooperate} is the most suitable strategy for $\alpha$. These results are in line with those obtained for random-action profiles with respect to equality (Fig. \ref{fig:random_pd_algn_eq}), where we observed that the alignment equilibrium was reached when both payers behaved similarly.\par

In order to find the alignment equilibrium from results presented as in Fig. \ref{fig:alpha_strat_algn_eq}, we must look for the position(s) such that: (a) For constant cooperation probability of $\beta$, \textit{i.e.} by fixing the position along the $x$ axis, changing $\alpha$'s strategy by switching line colour leads to a decrease in alignment; and (b) for fixed $\alpha$'s strategy, \textit{i.e.} maintaining the line colour, the cooperation probability of $\beta$ (the coordinate along the $x$ axis) corresponds to a maximum along that line.\par

For heterogeneous profiles, the alignment equilibrium with respect to equality is found when $\beta$'s strategy along the iterated game settles in cooperation (probability of cooperation equals 1) and $\alpha$ applies the \textit{tit-for-tat} or \textit{mostly cooperate} strategies, both of which result in $\alpha$ cooperating for all rounds of the game, except possibly in the first one. Neither $\beta$ nor $\alpha$ would then have any incentive to unilaterally deviate.\par

Differently to the case of random strategies, where there was an infinite number of alignment equilibria (the positive diagonal), in this case there is a single alignment equilibrium that corresponds to the persistent collaboration of both agents along the iterated game, which is something clearly desirable from a social perspective. Then, the introduction of strategic directives for $\alpha$ has shifted the equilibrium with respect to equality towards cooperation. It is also worth noting that, in this case, the equilibrium strategy corresponds exactly with the only Pareto optimal position.\par

Second, we concentrate on alignment with respect to personal gain, Fig. \ref{fig:alpha_strat_algn_gain}. In this case, $\alpha$'s alignment is strongly dependent on the cooperation probability of $\beta$. The three strategies yield alignments for $\alpha$ that are monotonically increasing with $\beta$'s collaboration. Also, the three strategies are always equally ranked. This means that, given a fixed cooperation ratio for $\beta$, the most aligned strategy is always \textit{mostly defect}, followed by \textit{tit-for-tat} and finally \textit{mostly cooperate}. The strategies are ordered from least to most cooperative.\par

As for $\beta$'s alignment, it increases linearly with its cooperation probability when $\alpha$ follows \textit{tit-for-tat}. It displays a peak at low collaboration rates when $\alpha$ deploys \textit{mostly cooperates}, and a valley at high cooperation probabilities when $\alpha$ follows \textit{mostly defect}.\par

Again, considering only the heterogeneous profiles that have generated these results, there are two alignment equilibria with respect to personal gain corresponding to $\beta$ not cooperating at all and $\alpha$ following either \textit{mostly defect} or \textit{tit-for-tat}. To achieve this conclusion, we first note that, for any cooperation probability of $\beta$, $\alpha$ always enhances her alignment by following \textit{mostly defect}. Then, once $\alpha$ has settled for this strategy, the best choice for $\beta$ is to always defect. These two observations lead to $\alpha$ following either \textit{mostly defect} or \textit{tit-for-tat} and $\beta$ never cooperating. Given that $\beta$ always defects, both these strategies converge to $\alpha$ always defecting as well, except maybe at the first round. It is worth noting that these equilibria are actually far from the maximum possible alignment for either agent. Nor do they result in a Pareto optimal alignment, since both agents could improve their alignment by having $\alpha$ follow \textit{tit-for-tat} and $\beta$ increase her probability of cooperation.\par

Finally, in an exercise similar to that performed for random-action profiles, we find the alignment equilibrium position when agents prioritise different values. Since players in this strategy profiles are not equivalent ($\alpha$ behaves in a conscious way while $\beta$ behaves completely randomly), we must examine two possibilities: $\alpha$ prioritises {\em personal gain} while $\beta$ prioritises {\em equality}, and vice-versa.\par

For the first possibility, $\alpha$ will always need to follow \textit{mostly defect} to ensure that her alignment with respect to personal gain is maximised, regardless of $\beta$'s probability of cooperation. Then, in order to attain maximum alignment with respect to equality, $\beta$ will settle on constant defection. Hence, the alignment equilibrium when $\alpha$ prioritises personal gain and $\beta$ prioritises equality is the same as when both prioritised personal gain. Yet again, these strategies do not lead to a Pareto optimal alignment, since both agents could improve the alignment with respect to their prioritised values by having $\alpha$ follow \textit{tit-for-tat} and $\beta$ increase her probability of cooperation.\par

For the second possibility, $\alpha$ will follow {\em tit-for-tat}, since this strategy dominates the two others when it comes to equality, regardless of the cooperation probability of $\beta$. Given this observation, $\beta$ will then enhance her alignment with respect to {\em personal gain} by always cooperating. At this position, $\alpha$ can resort to {\em tit-for-tat} or {\em mostly cooperate} indistinctly. This strategy profile is equal to the equilibrium found when both agents prioritised equality. In this case, it also corresponds to a Pareto optimal alignment, since $\alpha$ has achieved the maximum possible alignment with respect to equality.\par

It is worth pointing out that the alignment equilibrium positions under agents prioritising different values are driven by the player following the more conscious strategy, $\alpha$ in this case. That is to say that when $\alpha$ focuses on personal gain, the solution concepts are identical regardless of the value that $\beta$ (the randomly behaving agent) holds in high regard. The same result is found when $\alpha$ focuses on equality instead.\par

A summary with the results analysed for the model under heterogeneous profiles is provided in Table \ref{tab:heterogeneous_profiles_summary}.

\begin{table}[t]
    \centering
    \caption{Summary of the results obtained under heterogeneous profiles, highlighting the position of alignment equilibrium strategies and their Pareto optimality. (Nomenclature for strategies: TfT: {\em tit-for-tat}; MC: {\em mostly cooperate}; MD: {\em mostly defect}).}
    \begin{tabular}{|c|c|c|c|}
        \hline \textbf{Value for $\mathbf{\alpha}$} & \textbf{Value for $\mathbf{\beta}$} & \textbf{Alignment equilibrium} & \textbf{Pareto optimal} \\
        \hline Equality & Equality & $e_\alpha^*=\text{TfT/MC};\;e_\beta^*=1$ & Yes \\
        \hline Personal gain & Personal gain & $e_\alpha^*=\text{TfT/MD};\;e_\beta^*=0$ & No \\
        \hline Personal gain & Equality & $e_\alpha^*=\text{TfT/MD};\;e_\beta^*=0$ & No \\
        \hline Equality & Personal gain & $e_\alpha^*=\text{TfT/MC};\;e_\beta^*=1$ & Yes \\
        \hline
    \end{tabular}
    \label{tab:heterogeneous_profiles_summary}
\end{table}

\section{Conclusions and future work}\label{sec:conclusions}
In this work, we have reviewed  a formal framework that establishes preferences over the states in the world, and we have specified the computation of value alignment through the increase or decrease in preferences over states in a normative world. Some further study needs to be done in the theoretical front, but for the time being we have been able to implement this framework in the Iterated Prisoner's Dilemma model.\par

Inspired by the classical Nash equilibrium and Pareto optimality in game theory, we have introduced the novel notions of \textit{alignment equilibrium} and \textit{Pareto optimal alignment}. These solution concepts extend the existing definitions to account for different values beyond individual rewards and generalise to cases where agents may have different priorities. We have been able to identify both equilibria and Pareto optimal alignments in our Two-Agent Iterated Prisoner's Dilemma model, with respect to the values {\em equality} and {\em personal gain}. An interesting finding is that, under both strategy subsets under consideration, alignment equilibria positions with respect to equality include Pareto optimal outcomes, while equilibrium positions with respect to value gain do not.\par

This work intends to exemplify an application of the proposed model for the value alignment problem. Further work, built on it, remains to be done. First, on the analytical side, formal properties of the alignment equilibrium and its relationship with Pareto optimal alignments should be explored. Second, the model can be naturally extended to account for the introduction of norms, such as taxes, fines or the banning/enforcement of behaviour. An interesting outcome from such research should be the shift, if any, in the alignment equilibria positions. Finally, a third line of work should be focused on the development of methodologies to synthesise norms with optimal alignment with respect to values of choice.


\section*{Acknowledgments}
This work has been supported by the AppPhil project (RecerCaixa 2017), the CIMBVAL project (funded by the Spanish government, project \# TIN2017-89758-R), the EU WeNet project (H2020 FET Proactive project \# 823783) and the EU TAILOR project (H2020 \# 952215).

\bibliographystyle{splncs04}
\bibliography{mybibliography}

\end{document}